\documentclass[sigconf]{acmart}
\pdfoutput=1

\usepackage{amsmath,amssymb,amsfonts}
\usepackage{graphicx}
\usepackage{textcomp}
\usepackage{xcolor}
\usepackage{algorithm}
\usepackage{algpseudocode}
\usepackage{caption}
\usepackage{subcaption}

\AtBeginDocument{%
  \providecommand\BibTeX{{%
    \normalfont B\kern-0.5em{\scshape i\kern-0.25em b}\kern-0.8em\TeX}}}

\copyrightyear{2020}
\acmYear{2020}
\setcopyright{acmcopyright}\acmConference[SEAMS '20]{IEEE/ACM 15th International Symposium on Software Engineering for Adaptive and Self-Managing Systems}{October 7--8, 2020}{Seoul, Republic of Korea}
\acmBooktitle{IEEE/ACM 15th International Symposium on Software Engineering for Adaptive and Self-Managing Systems (SEAMS '20), October 7--8, 2020, Seoul, Republic of Korea}
\acmPrice{15.00}
\acmDOI{10.1145/3387939.3391595}
\acmISBN{978-1-4503-7962-5/20/05}

\begin{document}

\title{A Hybrid Approach Combining Control Theory and AI for Engineering Self-Adaptive Systems}

\author{Ricardo Diniz Caldas}
\affiliation{%
  \institution{Chalmers \textbar \, University of Gothenburg}
  \city{Gothenburg}
  \country{Sweden}
  \institution{\\University of Bras\'ilia, Brazil}
  }
\email{ricardo.caldas@chalmers.se}

\author{Arthur Rodrigues}
\affiliation{%
  \institution{University of Bras\'ilia}
  \city{Bras\'ilia, DF}
  \country{Brazil}}
\email{arthur.farias@aluno.unb.br}

 \author{Eric Bernd Gil}
 \affiliation{%
   \institution{University of Bras\'ilia}
   \city{Bras\'ilia, DF}
   \country{Brazil}}
 \email{ericbgil@gmail.com}

 \author{Gena\'ina Nunes Rodrigues}
 \affiliation{%
   \institution{University of Bras\'ilia}
   \city{Bras\'ilia, DF}
   \country{Brazil}}
 \email{genaina@unb.br}

\author{Thomas Vogel}
\affiliation{%
  \institution{Humboldt-Universit\"at zu Berlin}
  \city{Berlin}
  \country{Germany}}
\email{thomas.vogel@cs.hu-berlin.de}

\author{Patrizio Pelliccione}
\affiliation{%
  \institution{Chalmers \textbar \, University of Gothenburg}
  \country{Sweden}
  \institution{\\University of L'Aquila, Italy}}
\email{patrizio@chalmers.se}

\renewcommand{\shortauthors}{RD. Caldas, A. Rodrigues, EB. Gil, G. Rodrigues, T. Vogel, and P. Pelliccione}

\begin{abstract}
Control theoretical techniques have been successfully adopted as methods for self-adaptive systems design to provide formal guarantees about the effectiveness and robustness of adaptation mechanisms. However, the computational effort to obtain guarantees poses severe constraints when it comes to dynamic adaptation. In order to solve these limitations, in this paper, we propose a hybrid approach combining software engineering, control theory, and AI to design for software self-adaptation. Our solution proposes a hierarchical and dynamic system manager with performance tuning. Due to the gap between high-level requirements specification and the internal knob behavior of the managed system, a hierarchically composed components architecture seek the separation of concerns towards a dynamic solution. Therefore, a two-layered adaptive manager was designed to satisfy the software requirements with parameters optimization through regression analysis and evolutionary meta-heuristic. The optimization relies on the collection and processing of performance, effectiveness, and robustness metrics w.r.t control theoretical metrics at the offline and online stages. We evaluate our work with a prototype of the Body Sensor Network (BSN) in the healthcare domain, which is largely used as a demonstrator by the community. The BSN was implemented under the Robot Operating System (ROS) architecture, and concerns about the system dependability are taken as adaptation goals. Our results reinforce the necessity of performing well on such a safety-critical domain and contribute with substantial evidence on how hybrid approaches that combine control and AI-based techniques for engineering self-adaptive systems can provide effective adaptation.
\end{abstract}


\maketitle

\section{Introduction}

\vspace{1em}
There is an increasing trend in the use of control theory to guide the software engineering design process of self-adaptive systems (SAS) in order to provide dynamic adaptation with guarantees of trustworthiness and robustness~\cite{weyns:2018, shevtsov:2018,filieri:2017}. Control theory has established solid techniques for designing controllers that enforce controlled (or managed) systems to behave as expected. ``These controllers can provide formal guarantees about their effectiveness, under precise assumptions on the operating conditions''~\cite{filieri:2017}.
 
At runtime, control systems are subject to inputs not known at design-time. Based on that, they are designed to respond within acceptable boundaries to dynamic environmental interactions. The analysis of whether the controlled system is operating accordingly demands metrics for comparing the performance of various control systems~\cite{ogata:2001}. 
A self-adaptive system designed through control theory must be able to quantitatively guarantee its convergence in time to reach the adaptation goal (i.e., setpoint), and on its robustness (i.e., convergence to the setpoint) in the face of errors and noise~\cite{shevtsov:2018,filieri:2017}. Additionally, due to high level of uncertainty in self-adaptive systems, the control model following a control-theoretical approach requires constant evolution. 

Therefore, in the control-theoretical design of a self-adaptive software system, it is paramount to support decision-making procedures that can identify appropriate adaptation strategies by efficiently exploring the solution space. To avoid the problem of exhaustive analysis of large adaptation spaces when planning an adaptation, artificial intelligence (AI) techniques have been used, among others: to determine at runtime subsets of adaptation options from a large adaptation space through classification and regression~\cite{quin:2019}; to assess and reason about black-box adaptation decisions through online learning~\cite{Esfahani:2013, Elkhodary:2010}; or to find Pareto-optimal configurations of the system with respect to the mission goals of mobile robots through machine learning~\cite{jamshidi:2019}. 
Moreover, control-based approaches for self-adaptive systems often employ adaptive controllers to address inaccuracies of the control model or radical changes of the environment/controlled system by updating the model using filters (e.g., Kalman or Recursive Least Square)~\cite{Maggio:2012,Maggio:2014,filieri2015software} and by tuning the controller parameters using feedback or machine learning~\cite{Filieri:2012,Lama:2013}.

Although such significant contributions in the literature have promoted solid mathematical foundation of control theory for building SAS, there is still the need of understanding (i)~how AI could aid the parameter optimization of controllers and (ii)~how to apply AI-based techniques in the search of adaptation solutions in control-theoretical SAS strategies. Moreover, ``there are various hurdles that need to be tackled to turn control theory into the foundation of self-adaptation of software systems''~\cite{weyns2017software}. Software engineers are not usually knowledgeable of the mathematical principles on control theory. Therefore, there is a need to turn control theory (and its guarantees) into a scientific foundation for engineering self-adaptation~\cite{weyns2017software}.

In this paper, we aim at filling those gaps by intertwining control theory and AI in a two-phase optimization approach to support the engineering of control-based SAS. Our controller, hereafter named system manager, contains the mechanisms for deciding how the self-adaptive system must behave in order to achieve the desired goals for the managed system. Due to the gap between high-level requirements specification and the internal knob behavior of the managed system, a hierarchically composed components architecture seek the separation of concerns towards a dynamic solution~\cite{braberman:2015}. Therefore, our system manager is further decomposed into two other components: a strategy manager and a strategy enactor. Each of these components are equipped with AI-based optimization techniques, namely regression model combined with the Non-dominated Sorting Genetic Algorithm II (NSGA-II)~\cite{deb2002fast}, for respectively: (i)~synthesizing adaptation strategies through high-level reasoning upon the model that represents the managed system, and (ii)~enforcing actions through control-theoretical principles to ensure the adaptation strategies are applied and properties of interest guaranteed. By these means, we contribute with a hierarchical and adaptive system manager that relies on  optimization at all levels of the decision-making process towards a more efficient adaptation mechanism, while also improving the self-adaptation loop in terms of control theoretic properties: stability, overshooting and steady-state error.

For evaluation purposes, we experimented with a healthcare domain prototype constructed for operating volatile environments. The prototype played the role of the target system. The Body Sensor Network~(BSN) prototype was implemented following the Robot Operating System~(ROS) framework and consists of a set of distributed sensors that collects the patients vital signs and forwards them to a centralized unit for further processing and reporting. We evaluated the ability of our approach to optimize the adaptation space and to improve the self-adaptation loop w.r.t. the control theoretic properties. Results show that we have found nearly optimal solutions for space and time through our hybrid approach. Not to mention, the overshoot and steady-state error were below 3\% with 100\% of stability.

The remainder of this paper is organized as follows: Section \ref{sec:background} presents a glimpse over control theoretical principles and over evolutionary algorithms used in this work. Section \ref{sec:proposal} presents the core contribution of our work. Section \ref{sec:experiment} presents the evaluation of our proposal on the BSN prototype. Section \ref{sec:related} presents the most related literature work to ours. Finally, Section \ref{sec:conclusion} presents the conclusion and the directions we intend to pursue as our future work.

\section{Background}\label{sec:background}

In this section, we discuss the background of our work, feedback control loop and evolutionary algorithms.

\begin{figure}[htbp]
    \centerline{\includegraphics[scale=0.135]{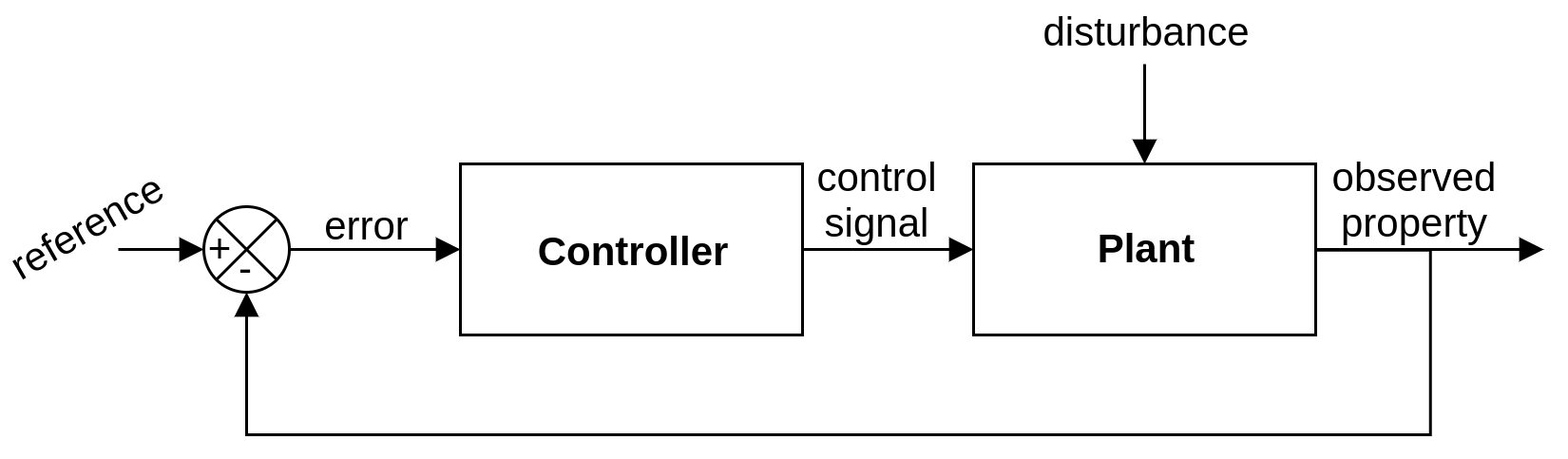}}
    \caption{Block diagram of a closed feedback control loop.}
    \label{Feedback Loop}
\end{figure}

\noindent\textbf{Feedback Control Loop.}
Self-adaptation is typically realized by a feedback loop~\cite{Brun:2009}, which originates from control engineering~\cite{Astrom:2010}. As shown in Fig.~\ref{Feedback Loop}, the \textit{plant} is the subsystem to be controlled by a control signal from the \textit{controller}. 
A reference (setpoint) is defined for a property of interest observed on the plant. The goal is that the observed property is sufficiently close to this reference despite the disturbance affecting the plant.
For this purpose, the difference between the observation and reference is fed back to the controller that uses this error to decide about the value of the control signal.

\begin{figure}[b]
    \centering
    \centerline{\includegraphics[scale=0.16]{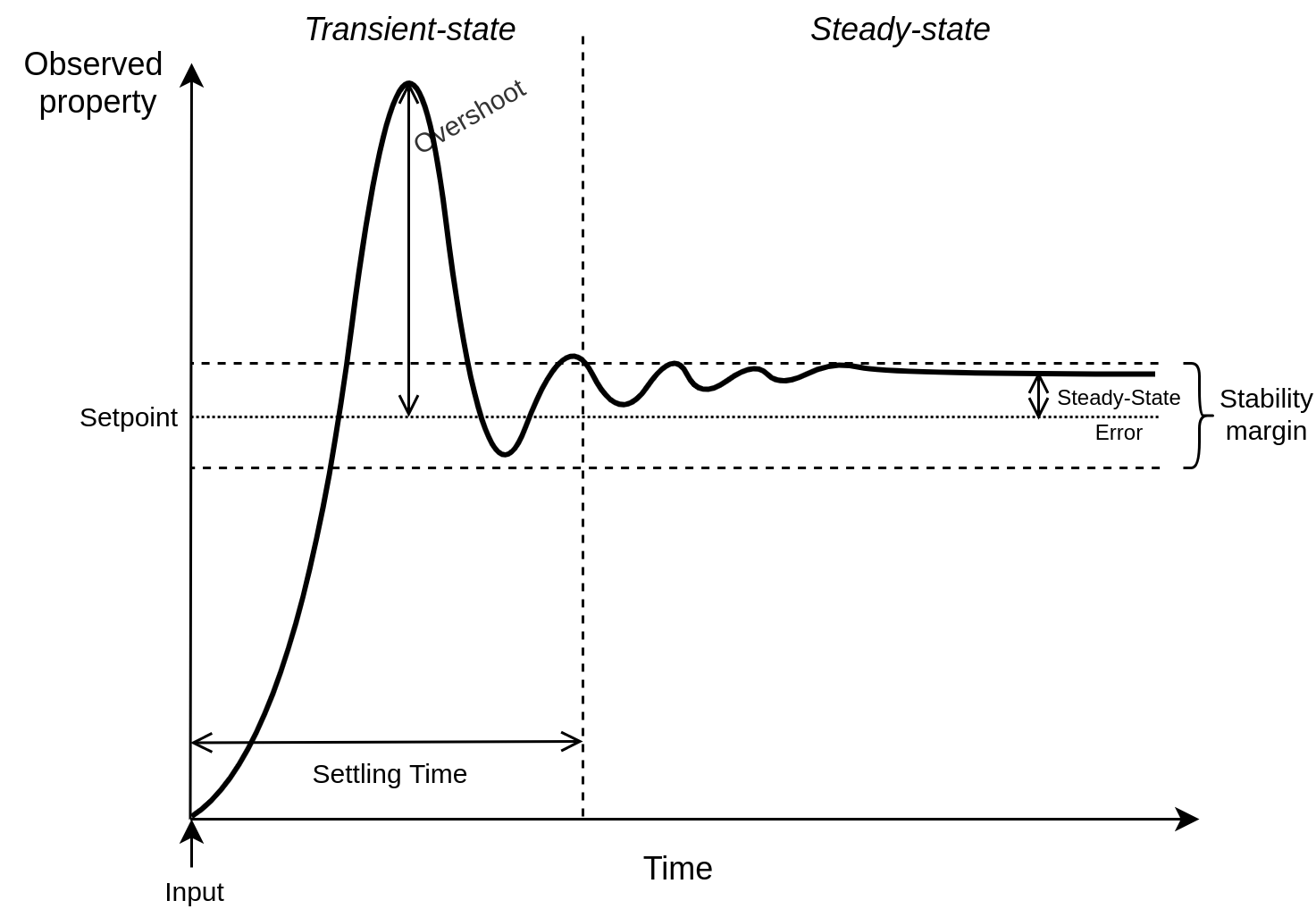}}
    \caption{Typical response of a system to a sudden variation on the observed property.}
    \label{fig:resp-w-metrics}
\end{figure}

A self-adaptive system that uses a feedback control loop to adjust its behavior is a closed-loop system~\cite{Hellerstein:2004,Salehie:2009}.
Such a system can have different responses to a perturbation, that is, how the property of interest observed on the plant develops over time after the perturbation. Figure~\ref{fig:resp-w-metrics} illustrates a typical response to a sudden variation in the observed property. 
Various control metrics exist to verify how well the system is responding and if it satisfies requirements previously defined with expert knowledge (cf.~\cite{filieri:2017,Hellerstein:2004}).
In this work we focus on extracting four metrics to analyze the response of the system:
\emph{stability},
\emph{overshoot},
\emph{settling time}, and 
\emph{steady-state error}~\cite{filieri:2017,Hellerstein:2004}. 

Stability refers to the system reaching a steady-state equilibrium, that is, the observed property converges to a specific value, ideally the setpoint, and stays inside a previously defined stability margin around this convergence point.  
An overshoot denotes that the property exceeds the setpoint in the transient state, which should be typically avoided. It is measured as maximum peak value of the property relative to the setpoint.
The settling time is the time required for the system to reach the steady-state equilibrium. 
Finally, the steady-state error measures how far from the setpoint the system converges, that is, the steady-state difference between the setpoint and observed property.

In a feedback loop, one of the most popular controllers being used is the proportional-integral-derivative (PID) controller while often the derivative part is omitted due to difficulties in its tuning, leading to the use of PI controllers (cf.~\cite{ang2005pid}).
Moreover, such controllers are also widely used for adapting software systems~\cite{shevtsov:2018}.
Based on this observation, we also use a PI controller throughout this work.
The continuous equation of a PI controller is shown in Equation~\ref{eq:PI}, where $u(t)$ is its control action and $e(t)$ the error, both at time $t$. In this equation we have three constants, which are the proportional gain $K_{p}$, the integral gain $K_{i}$. and the integral window $IW$. $K_{p}$ multiplies the current measured error directly, whereas $K_{i}$ multiplies the integral of the error over the time window $IW$. 
These constants can be defined empirically or using a tuning method of preference. 

\begin{equation}
    \label{eq:PI}
    u(t) = K_{p}e(t) + K_{i}\int_{j=t-IW+1}^{t} e(j) 
\end{equation}

A discrete approximation of Equation~\ref{eq:PI} is shown in Equation~\ref{eq:PIdiscrete}, which is adapted from the work of \citet{stankovic1999case}.

\begin{equation}
    \label{eq:PIdiscrete}
    u(t) = K_{p}e(t) + K_{i}\sum_{j=t-IW+1}^{t} e(j) 
\end{equation}

In this equation, the integral term is transformed into a sum of errors over the time window $IW$.  
Since a PI controller is also called periodically, the time window $IW$ is a tunable parameter, which we consider in this work as the number of error measurements to be considered in the sum. This means that in the time instant $t$ the errors from $e(t-IW+1)$ to $e(t)$ will be used in the integral term.

~\noindent\textbf{Evolutionary Algorithms.} 
Evolutionary algorithms originate from computational intelligence, a subfield of artificial intelligence, and refer to meta-heuristic optimization techniques that imitate biological evolution~\cite{Kruse:2016}.
A widely-used technique is the \textit{Non-dominated Sorting Genetic Algorithm II (NSGA-II)}~\cite{deb2002fast} that has been applied to various software engineering problems~\cite{Harman:2012}, and self-adaptive systems to optimize the configuration of the managed system~\cite{Fredericks:2019,Shin:2019}.
NSGA-II is a multi-objective algorithm producing a set of Pareto-optimal solutions to a given optimization problem. Imitating biological evolution, it evolves a population of possible solutions encoded as chromosomes by crossover, mutation, and selection. The evolution is guided by a fitness function that encodes the objectives of the optimization and that evaluates how well a solution satisfies these objectives. Solutions with a higher fitness are preferably selected for further evolution steps in the next generation of the population.
Thus, the evolution converges to solutions with higher fitness.
Moreover, NSGA-II promotes diversity to obtain Pareto-optimal solutions with different trade-offs between the~objectives.

\section{A Hybrid Approach Combining Control Theory and AI} \label{sec:proposal}

\vspace{1em}

In this paper, we propose an AI-based optimization approach to support the engineering of control-based SAS. 
Following principles of control theory, our system manager decides how the system must behave in order to achieve the desired goals for the managed system. Moreover, our system manager is enhanced with integrated AI-based optimization techniques, namely regression  at all levels of the decision-making process towards a more efficient adaptation mechanism for (i)~high-level reasoning upon the model that represents the managed system to synthesize adaptation strategies (\emph{cf.} Section \ref{subsec:Manager}) and (ii)~enforcing actions through control-theoretical principles ensuring the adaptation strategies respect the properties of interest: stability, overshooting and steady-state error (\emph{cf. Section \ref{subsec:Enactor}).} 
In our approach, an adaptation strategy is a rule that guides the system behavior. It consists of a goal to be reached, a condition that defines whether the goal is reached, and one or more actions to change the system behavior.

\begin{figure}
    \centering
    \centerline{\includegraphics[scale=0.12]{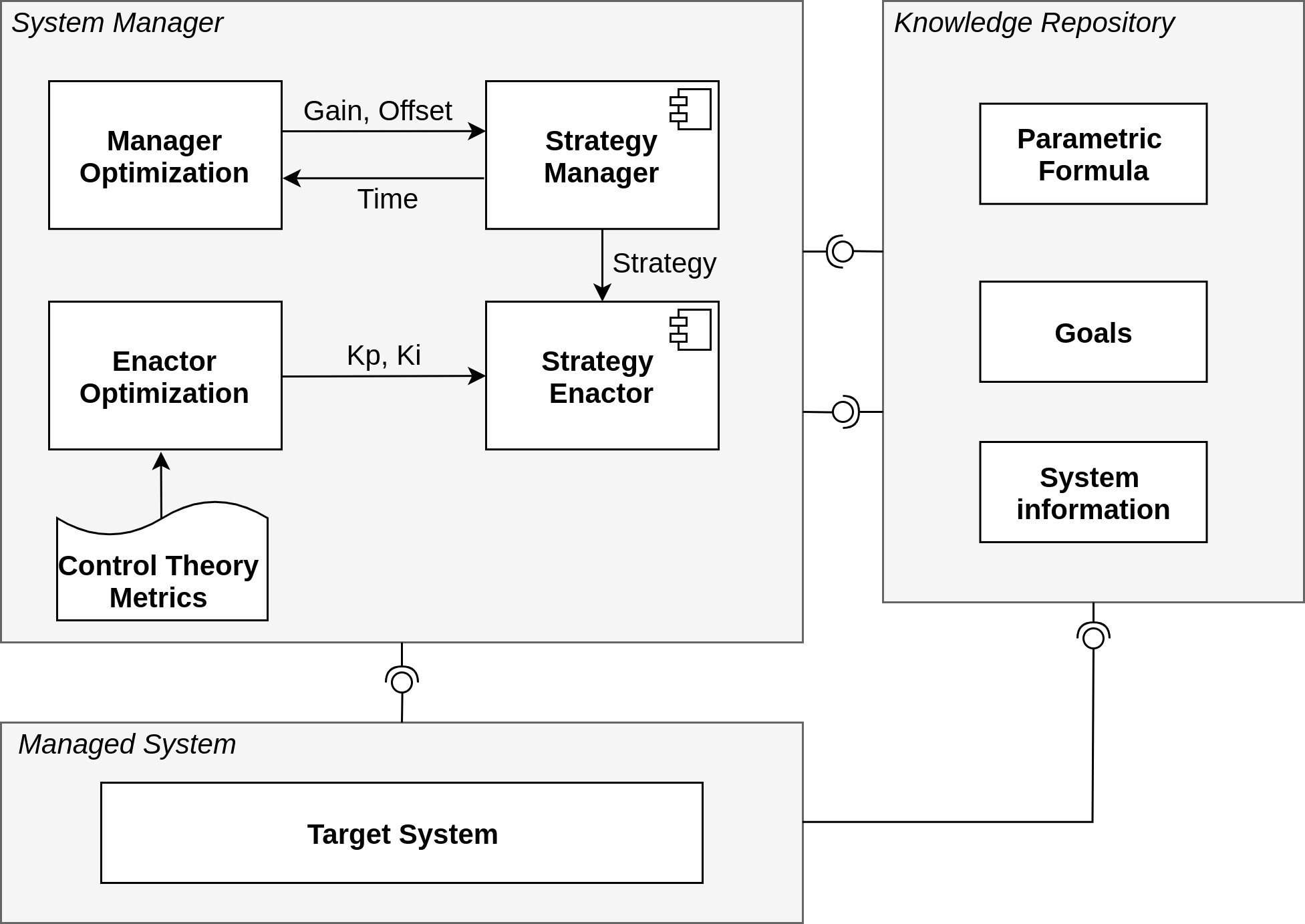}}
    \caption{An architectural overview of the hybrid approach.}
    \label{fig:process}
\end{figure}

To provide a solution that architecturally promotes the separation of concerns in a hierarchical fashion~\cite{braberman:2015}, the proposed architecture relies on two horizontal layers being  the \textit{System Manager} and the \textit{Managed System}, and a third and vertical layer being the \textit{Knowledge Repository}. The corresponding architectural overview of our approach is shown in Fig.~\ref{fig:process}.

The \textit{Managed System} layer contains the \textit{Target System} being the software system to be controlled.
The \textit{Knowledge Repository} is orthogonal to the other layers and responsible for maintaining the persistent information, that is, a \textit{parametric formula}, the \textit{goals}, and \textit{system information}. 
A parametric formula evaluates a quality of service (QoS) property of the target system at runtime. For details of how such a formula is created, we refer to our previous work~\cite{solano2019taming}.
The goals are the objectives to be achieved by the system with respect to functional or non-functional requirements. 
The system information is a set of events and data about the status of the target system  collected from the running system.

The \textit{System Manager} is responsible for controlling the \textit{Target System} and realizes the combination of control-based self-adaptation and artificial intelligence (AI), which makes our approach hybrid and improves the self-adaptation process.

For this purpose, the system manager has four components, two realizing self-adaptation in a hierarchical fashion and two realizing AI-based optimization to reconfigure the self-adaptation process. This results in a hierarchical and adaptive system manager.

The \textit{Strategy Manager} synthesizes adaptation strategies by reasoning on the target system using the parametric formula, the goals defined by the stakeholders and the current status of the observed property. 
This synthesis process and especially the time needed for the synthesis of adaptation strategies is monitored by the \textit{Manager Optimization} component. Based on the measured time, this component optimizes the configuration of the strategy manager by providing optimal values for two parameters (\textit{Granularity} and \textit{Offset}) that influence the synthesis process. The goal of the optimization is to reduce the required time and thus, the efficiency of the synthesis process. The strategy manager provides a synthesized strategy to the \textit{Strategy Enactor}.  

The strategy enactor enforces an adaptation strategy in a closed-loop fashion by implementing a feedback control loop on top of the target system.
Accordingly, the strategy enactor implements a PI controller as an actuation mechanism for enforcing behavior at the target system, which is guided by conditions and goals.
This self-adaptation process is monitored by the \textit{Enactor Optimization} component, that computes various \textit{Control Theory Metrics} based on the \textit{System Information} collected from the running target system. Moreover, the component uses the metrics to tune the strategy enactor adaptation parameters. 
Upon the metrics and system information, the optimizer tunes the PI controller parameters ($K_p$ and $K_i$).
The goal of this optimization is to improve the adaptation quality with respect to the control theory metrics.

As a result of combining each, the strategy manager and the strategy enactor with an optimization, the goals of the system are guaranteed to be constantly achieved with optimal adaptation search time and adaptation quality, in a limited search space. 
Both optimizations are realized by AI-based techniques of model learning (regression / curve fitting) and meta-heuristic search (evolutionary algorithms), while the strategy enactor is based on control (PI controller). By this combination, the overall controller (i.e., the system manager) implements a hybrid approach of control and AI.

We evaluate our approach with the Body Sensor Network (BSN)\footnote{Access \url{http://bodysensornetwork.herokuapp.com} for an executable artifact.}, being an exemplar for a target system that monitors and analyzes the health of patients.
The BSN is composed of a set of distributed sensors to monitor vital signs of patients and to forward these signs to a centralized processor for analysis. To avoid overload of the centralized processor, the sampling rate of individual sensors can be adapted, which effects the amount of data sent to and analyzed by the processor.

In the following we detail the stategy manager and strategy enactor of our hybrid approach emphasizing their optimization.

\subsection{Strategy Manager and its Optimization}\label{subsec:Manager}

Responsible for high-level reasoning, the strategy manager must be capable of performing expensive decision-making computation. Identifying feasible and effective adaptation strategies is particularly difficult as the size of the solution space grows exponentially with the number of individual adaptation options. 
For instance, for architectural adaptation the reasoning has to take the variability of each component (e.g., alternative components or parameters of the component) and the composition of components into account to evaluate whether the overall architectural configuration satisfies the goals of the system.

In previous work~\cite{solano2019taming} we have developed a transformation framework for specifying requirements hierarchically in a goal model, translating the goal model to parametric formulae with symbolic model checking, and evaluating these formulae at runtime to express probabilities over the fulfillment of the goals by the running system.
Thus, these formulae provide guarantees for satisfying the goals while their evaluation is time efficient (constant computational time) since a costly model exploration as in typical model checking is avoided.
The goal modeling and the translation of the goal model to parametric formulae are implemented in the GODA-MDP tool\footnote{https://pistar-goda.herokuapp.com/}~\cite{solano2019taming}.
A parametric formula is essentially an algebraic equation that relates the probability of successfully achieving a system's overall goal (i.e., the root goal of the goal model) to the combined probabilities of lower-level tasks that contribute towards realizing the goal. 
Therefore, a formula can be used in the analysis and planning stages of a feedback loop to solve a satisfiability problem, where an appropriate combination of successful local tasks would lead to satisfying a required global property. Searching for such a combination corresponds to the synthesis of adaptation strategies since the combination corresponds to a configuration that can be enacted on the target system.
Although using the parametric formula reduces the complexity and time for reasoning, such formulae are still not free from problem of combinatory state-space explosion.

\begin{algorithm}[!ht]
    \noindent\textbf{Input:} \textit{gran},\;\textit{offset} \\
    \textbf{Output:} \textit{strategy} 
    \begin{algorithmic}[1]
        \Procedure{Search For Strategy}{}
            \State $P \gets \textit{MonitorSystem}$
            \State $p_{curr} \gets Apply(P, model)$
            \State {$\textit{error} \gets p_{ref} - p_{curr}$}
            
            \For {$\textbf{each} \; p_{i} \; \textbf{in} \;  P$}

                \State {$P_{s} \gets \textit{ResetToStartingPoint(P, offset)}$}
                
                \State {$p_{new} \gets Apply(P_{s}, model)$}

                \If {$\textit{error} > 0$}

                    \State {\textbf{do} $p_{i} \gets p_{i} + \textit{gran}\cdot\textit{error} \; \textbf{until} \; p_{new} > p_{ref}$}

                    \State {$\textbf{where} \; p_{new} \gets Apply(P_{s}, model)$}

                    \For {$\textbf{each} \; P_{j} \; \textbf{in} \;  P_{s} - p_{i}$}

                        \State {\textbf{do} $p_{j} \gets p_{j} + \textit{gran}\cdot\textit{error} \; \textbf{until} \; p_{new} > p_{ref}$}

                        \State {$\textbf{where} \; p_{new} \gets Apply(P_{s}, model)$}

                    \EndFor
                \Else
                    \State {$p_{i} \gets p_{i} - gran\cdot\textit{error} \; \textbf{until} \; p_{new} < p_{ref}$}
                    
                    \State {$\textbf{where} \; p_{new} \gets Apply(P_{s}, model)$}

                    \For {$\textbf{each} \; p_{j} \; \textbf{in} \;  P_{s} - p_{i}$}

                        \State {\textbf{do} $p_{j} \gets p_{j} - \textit{gran}\cdot\textit{error} \; \textbf{until} \; p_{new} < p_{ref}$}

                        \State {$\textbf{where} \; p_{new} \gets Apply(P_{s}, model)$}

                    \EndFor
                \EndIf

                \State {$\textit{Strategies} \gets \textit{Strategies} + P_{s}$}

            \EndFor

            \State {$\textbf{return} \textit{ strategy from Strategies}$}

        \EndProcedure
    \end{algorithmic}
    \caption{- Strategy Manager Procedure}
    \label{alg:adpt_engine}
\end{algorithm}

The pseudocode of Algorithm~\ref{alg:adpt_engine} details the search process to solve a parametric formula captured in a \textit{model}. The search process has two input parameters, the \textit{granularity} and \textit{offset} that are used for reducing or widening the solution/search space.
Lines 2-4 capture the monitoring of a set of QoS properties $P$ of the target system, the calculation of the current QoS property $p_{curr}$, and the calculation of the $error$, that is, how far from the setpoint $p_{ref}$ the current property value $p_{curr}$ is. 
Then, lines 5-24 represent the search for a combination of independent terms ($p_{x}$), for each $p_{i}$ in the set of all properties $P$, that would lead to the goal ($p_{ref}$). 
To do so, lines 6-7 reset all $p_{x}$ into the value defined by the $offset$ and the set of new values ($P_s$) is plugged into the model. If the error turns to be positive, then the value of each of the properties in $P_s$ should be incremented with the error multiplied by a factor dictated by the granularity $gran$ until they, one by one, reach values that approximate the dependent value to the goal (see lines 8-15). On the other hand, the properties in $P_s$ are decremented by the same factor (lines 15-22). At line 23, the modified set of $P_s$ is concatenated into the valid set of \textit{Strategies}. In the end, the best or a sufficient strategy is chosen among the valid ones (line 25).

In this search process, the procedure \textit{Apply} matches the independent variables of the model/formula to each respective value, and performs the calculation that results in a value for the dependent variable, that is, the procedure evaluates the formula. Thus, it is used for quantitative reasoning on the system state. As a result, the algorithm relies on performing a search within the combination of all independent variables that would lead the system into reaching the goal. The search itself increases/decreases the value of each independent variable at a time, by gradually changing its value with a step dictated by a factor of $\textit{granularity}\cdot\textit{error}$. 

The granularity and the offset parameters are fundamental to determine whether the strategy manager will be able to find a combination of values (i.e., to converge to a solution) that satisfies the goal of the system, and to find such a combination in time.
The smaller the granularity is, the algorithm will more likely find a solution, even though it would take more time to do so. In contrast, the offset broaden the search space boundaries, so the bigger it is the more values will the search go through until it reaches the end line.
Consequently, we need a method to enhance the choice of such parameters, and this is where we employ AI-based optimization.\\

\noindent\textbf{Manager Optimization.}
In the following, we discuss the optimization of the search process conducted by the strategy manager to synthesize adaptation strategies. The optimization aims for identifying suitable values for the \textit{granularity} and \textit{offset} parameters (cf.~Algorithm~\ref{alg:adpt_engine}).
The optimization of the strategy manager is a single-objective parameter optimization problem since the goal is to minimize the time required by the strategy manager to synthesize an adaptation strategy. This problem is solved by the optimization pipeline depicted in Figure~\ref{fig:opt_strman}.
This pipeline consists of the following four steps:

\vspace{-1em}
\begin{figure}[h]
 \centering
 \centerline{\includegraphics[width=1\columnwidth]{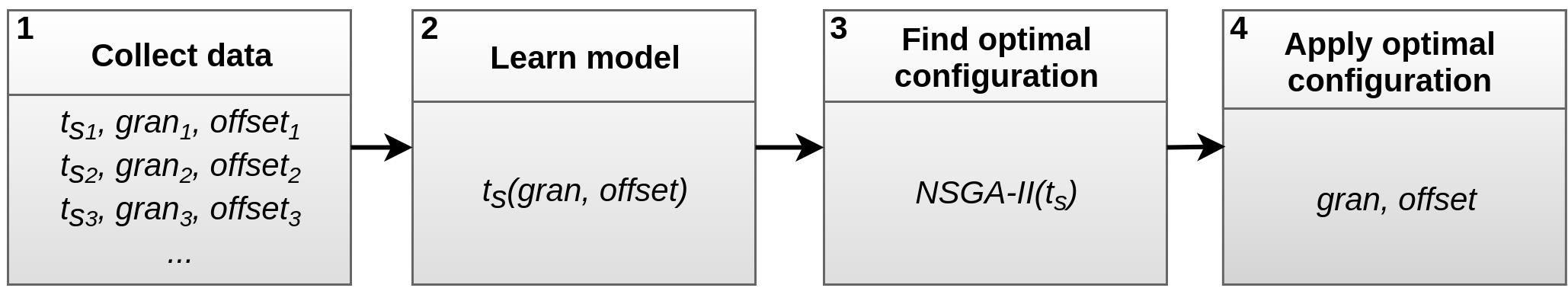}}
    \caption{Optimization pipeline for strategy manager.}
 \label{fig:opt_strman}
\end{figure}

\noindent \textbf{1. Collect Data:} 
We collect data relating the time to find a solution ($t_{s}$) with values for the granularity ($gran$) and \textit{offset} parameters. The variable $t_{s}$ represents the time taken by the strategy manager to converge to the setpoint. The data is obtained from executing our approach without the optimization in place. Thus, we execute our approach in different scenarios, that is, with different static choices of parameter values for granularity and offset. 

\noindent \textbf{2. Learn Model:} From the collected data <\textit{time, gran, offset}>, we learn a model using the curve fitting method. The learned model describes the behavior of the strategy manager as a function $t_{s}(\textit{gran, offset})$. Thus, we can use the model to estimate the time to synthesize an adaptation strategy given concrete values for the granularity and offset parameters. An example for such a model is $time = 677*e^{gran*(-3000)}+e^{offset*(-2874)}$. 

\noindent \textbf{3. Find Optimal Configuration:} We use the learned model as an input to an evolutionary algorithm, particularly NSGA-II, to find optimal values for the granularity and offset parameters in order to minimize the time to synthesize an adaptation strategy. 
For the evolutionary algorithm, a candidate solution is encoded as a chromosome or a vector two components, one component for each parameter. The fitness function is the minimization of the time. The fitness of a specific candidate solution, that is, a concrete pair of two real values for the two parameters, corresponds to computing the learned function with this pair of values.
Moreover, the optimization is constrained by the problem domain, as well as the range of the granularity and offset parameters. The optimization performed by evolutionary algorithm results in an optimal or near-optimal configuration (parameter setting) for the strategy manager. To illustrate, lets bring back our previous example model, $time = 677*e^{gran*(-3000)}+e^{offset*(-2874)}$. Initially, a population of one hundred individuals is generated randomly, having the target values and/or input parameters limited by a given range. An individual is composed of a combination of genes and a calculated target value. In NSGA-II, the genes are represented by the input parameters, in this case \textit{gran} and \textit{offset}. The target value in this step is the $time$ variable. Better ranked solutions, i.e., solutions with the best fitness values, are observed in individuals with a lower $time$ to find a solution. For the optimization of both, strategy manager and strategy enactor stages, we have set the algorithm to process a hundred generations. Moreover, we have applied some standard operators like Probability Matching (PM), Simulated Binary Crossover (SBX), and Tournament Selector based on dominance comparison (Pareto).

\noindent \textbf{4. Apply optimal configuration:} Finally, we apply the optimal or near-optimal configuration, that is, a pair of values (\textit{gran, offset}), to the strategy manager affecting the performance of the synthesis of adaptation strategies.

Delving into the optimization pipeline, we start describing the first part of the \textit{learning}\footnote{Strictly speaking, in this stage the optimization modules do rely on previous adaptation data to learn the behavior and optimize the strategy manager parameters. However, we reckon that the use of term \textit{learning} to describe curve fitting methods and evolutionary algorithms is still disputable.} process: the \textit{manager optimization}. As mentioned before, the manager behavior is ruled by two features: (i) \textit{gran}, that is the granularity with which the strategy manager will search for new solutions, and (ii) \textit{offset}, representing the starting point of the search within the solution space. Both attributes are crucial to determine a third variable of interest, the (iii) \textit{time to solution}, i.e., how long the manager will take to find a suitable adaptation strategy. From a naive perspective, these values are often defined in an uninformed fashion, before the system operation, and remain unchanged despite the characteristics of the system, or the particularities of the domain. Our proposal tweaks this logic by tracing the strategy manager behavior from its data collected offline.
Knowing the fact that the target variable depends on the \textit{gran} and \textit{offset} values, we conduct a curve fitting process to come up with models that explain the \textit{time to solution} variable in terms of the aforementioned parameters. Such a process returns a mathematical function that has the best fit to the data points collected from previous executions of the strategy manager. 
If the learned function being the output of the curve fitting process is simple, it is possible to solve the optimization problem mathematically. 
Otherwise, we solve this problem using an evolutionary algorithm, like the NSGA-II.

\vspace{1em}

We should note that our approach has been conceived to provide informed choices about granularity, offset, $K_p$ and $K_i$ at design time. Therefore, the amount of data used for training is not a major concern. A better description of the system behavior allows the engineers to make better choices of control theoretical metrics. It may be the case that the system behavior changes in face of uncertainty. Clearly, learning online from data and continuously feeding the system with new control theoretical metrics may leverage the strength of our approach. However, adopting an online learning approach brings additional complexity to the pipeline, which is out of scope in this work.

\subsection{Strategy Enactor and its Optimization}\label{subsec:Enactor}

The strategy enactor is responsible for enforcing an adaptation strategy on the target system. The strategies are synthesized by the strategy manager and must be read, interpreted, and enforced by the strategy enactor. Under the guidance of the active strategy, the enactor continually evaluates whether the target system behaves as it is intended to. Otherwise, it adapts the system towards achieving its goals. Whenever the enactor cannot enforce the active strategy, an exception is propagated to the strategy manager in an attempt to receive a new and more adequate adaptation strategy from the strategy manager.

Moreover, the strategy enactor realizes a negative feedback loop. In this sense, the enactor continuously monitors the system QoS properties, analyzes them, and checks whether its status is compliant to the goal demanded by the active strategy. 
As previously discussed, a strategy is defined by a \textbf{goal}, a \textbf{condition}, and a \textbf{set of actions} to achieve a desired configuration of the target system.

\begin{figure}
    \centering
    \centerline{\includegraphics[width=1\columnwidth]{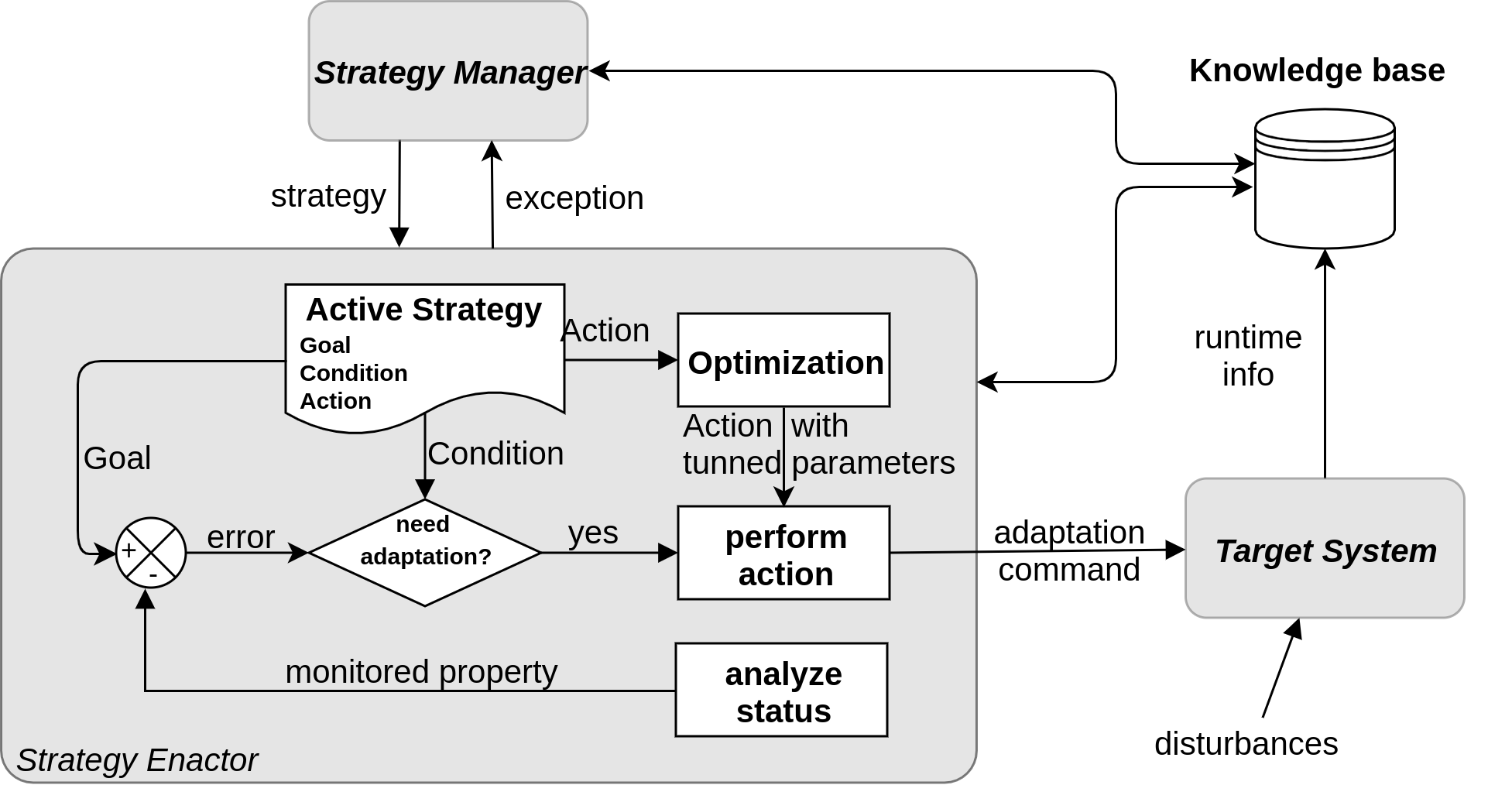}}
    \caption{Overview of the Strategy Enactor.}
    \label{fig:str_enact}
\end{figure}

Figure~\ref{fig:str_enact} depicts the configurable feedback loop for enforcing strategies on the target system. The strategy enactor collects information from the target system and analyzes the QoS property of interest. If the monitored property status does not correspond to the desired goal under the defined condition then the enactor enforces one or more actions. Whereas the actions are optimized accordingly.
In this context, we rely on control theory to implement the function that determines the control signal by using a PI controller. 
In this work, we use the BSN exemplar as a target system, whose knobs are the sampling rates of the sensors, which are adapted based on the reliability of the BSN's central processing component affected by the load on this component.

The synthesized adaptation strategy guides the parameters of the closed loop and how the adaptation is executed. To this extent, the PI controller is an instance of the element that enforces the control signal on the target system, characterized as the ``perform action'' in  Fig.~\ref{fig:str_enact}. Thus, it receives as input the difference between the current state of the observed system property and the goal of the adaptation strategy (error), and multiplies the error by factors of the gains Kp and Ki coming from the optimized action, according to Eq.~\ref{eq:PIdiscrete}. In the case of our running example, the PI controller would trigger an adaptation command with a description of the calculated new sampling rate for a sensor (e.g., the thermometer of the patient). Depending on the values of Kp and Ki, the increments or decrements on the actuation variable could not guide the property into converging to the setpoint, they could request too much effort from the system, or they could take too long to reach the setpoint. Foremost, the lack of information on a model that relates Kp, Ki, stability, overshoot and steady-state-error demands an optimization of Kp and Ki towards optimal performance. \\

\noindent\textbf{Enactor Optimization.}
The designed PI controller is not sufficient to guarantee optimal performance. Therefore, we employ an AI-based optimization technique to tune the controller's parameters with respect to the desired control-theoretical properties of \textit{settling time}, \textit{overshoot}, and \textit{steady-state error}. In this work, we focus on the \textit{overshoot} and \textit{steady-state error}.
These parameters are the proportional gain \textit{$K_p$} and the integral gain \textit{$K_i$} of the PI controller. 
Thus, the optimization problem is to find optimal values for \textit{$K_p$} and \textit{$K_i$} that improve the self-adaptation in terms of \textit{overshoot} and \textit{steady-state error}. These two control properties should be minimized resulting in a multi-objective optimization problem.

\begin{figure}[h]
 \centering
 \centerline{\includegraphics[width=1\columnwidth]{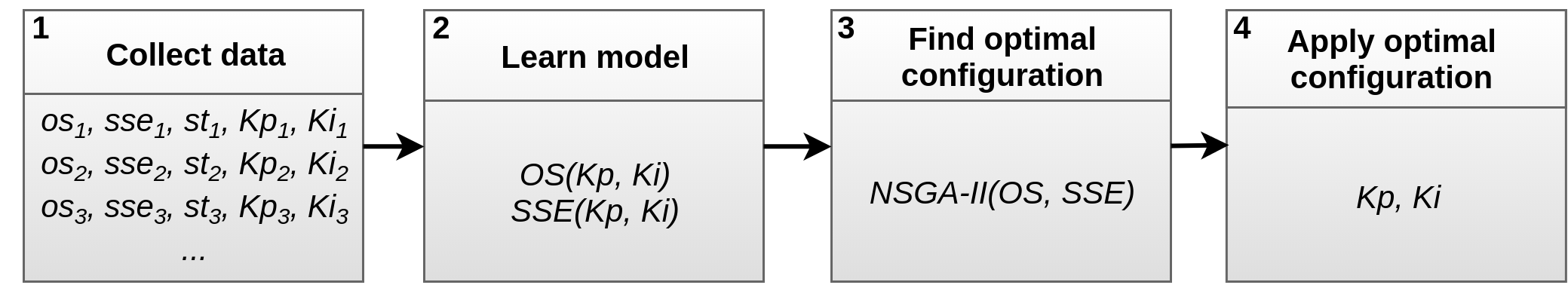}}
    \caption{Optimization pipeline for strategy enactor.}
 \label{fig:opt_strenc}
\end{figure}

\begin{figure}[h]
    \centering
    \centerline{\includegraphics[width=0.5\textwidth]{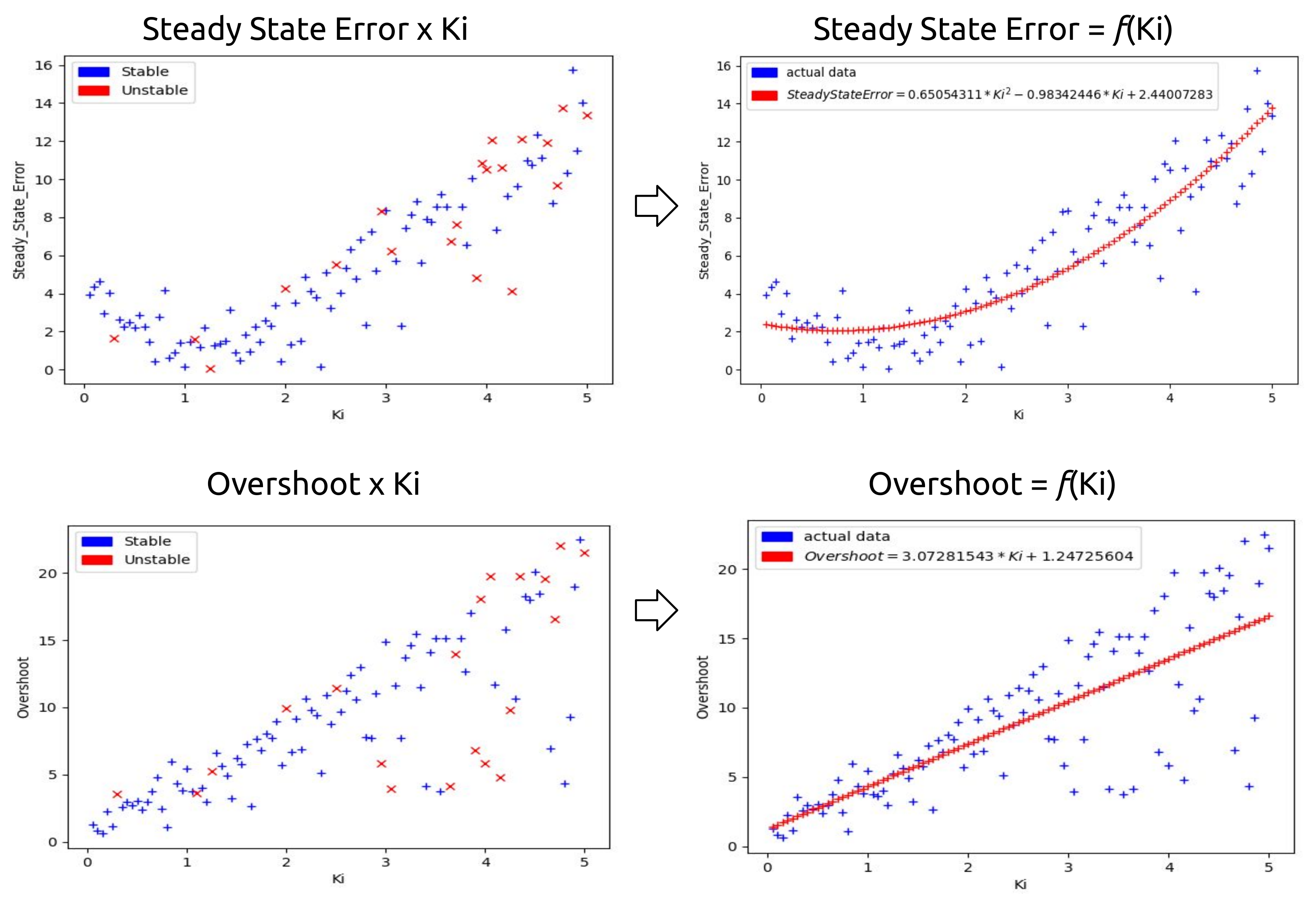}}
    \caption{Model learning by curve fitting the behavior of the strategy enactor.}
     \label{fig:curvefitting}
\end{figure}

The optimization pipeline for the strategy enactor is shown in Fig.~\ref{fig:opt_strenc}. Except for slight differences, this pipeline and the employed techniques are the same as for the strategy manager (cf.~Fig.~\ref{fig:opt_strman}).
Considering Fig.~\ref{fig:opt_strenc}, the control-theoretical metrics are collected from the behavior of the system during execution. From this data we get the data points relating $K_p$ and $K_i$ to settling time, overshoot, and steady-state error. Using this data, we adopt a curve fitting strategy to come up with three mathematical functions that describe the relationship between the parameters $K_p$ and $K_i$ and each control-theoretical metric: overshoot ($OS$) and steady-state error ($SSE$) (see second step in Fig.~\ref{fig:opt_strenc}). The type of the curve (e.g., linear, polynomial, exponential, etc.) depends on the relation revealed by the data points. This is illustrated in Fig.~\ref{fig:curvefitting} for the BSN case study.

\begin{figure}[h]
    \centering
    \centerline{\includegraphics[width=0.5\textwidth]{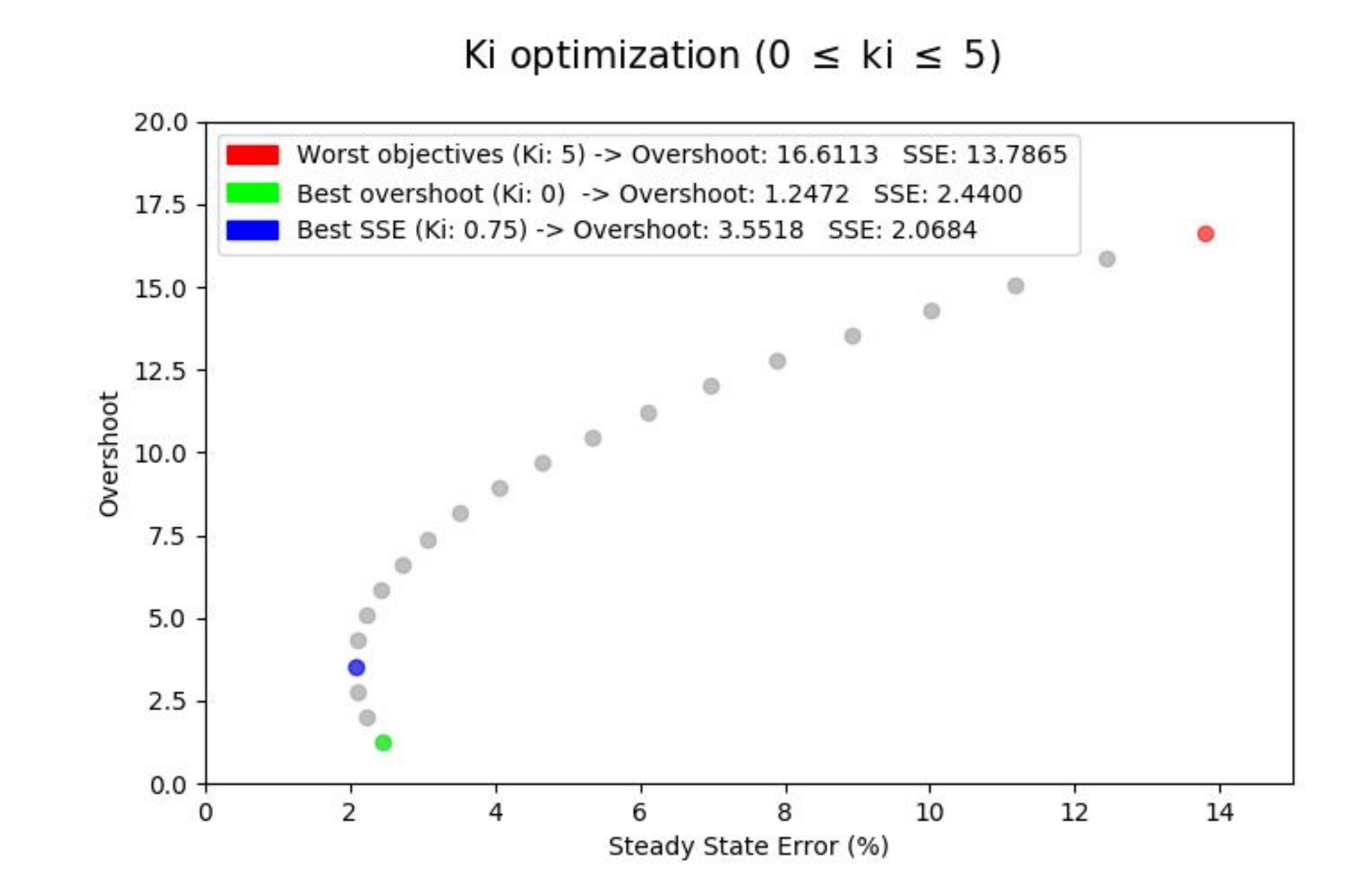}}
    \caption{Multi-objective optimization example}
    \label{fig:optimization}
\end{figure}

Once we have these functions, it is possible to feed the NSGA-II with them in order to find values of $K_p$ and $K_i$ that minimize the overshoot and steady-state error. Since we have now a multi-objective optimization problem, the meta-heuristic will not return just a single pair of optimal values for $K_p$ and $K_i$, but a set of Pareto-optimal solutions, i.e., several pairs of ($K_p, K_i$). These results are optimal or near-optimal solutions trading off the two objectives, overshoot and steady-state error as depicted in Fig.~\ref{fig:optimization}. 

Finally, we update the parameters of the PI controller employed in the strategy enactor with $K_p$ and $K_i$ values that optimize the gains that lead to lower and more balanced overshoot and steady-state error.

\section{Experimental Results} \label{sec:experiment}

Our proposal is evaluated through the execution\footnote{Configurations for experiments: CPU 2x Intel(R) Core(TM) i7-8550U CPU @ 1.80GHz, 8192MB RAM, Ubuntu 18.04.3 LTS, GNU C Compiler version 7.4.0} of an actual prototype from the healthcare domain (BSN) as the target system alongside an implementation\footnote{https://www.github.com/lesunb/bsn} of the system manager components upon ROS. The evaluation itself follows the Goal-Question-Metric methodology~\cite{caldiera:1994gqm}. 
Our focus is on answering the guiding questions, from a broader perspective (i)~how could AI aid the parameter optimization controllers, and (ii)~how to apply AI-based techniques in the search for adaptation solutions in control-theoretical SAS strategies. Thus, we narrow down the questions towards an operational scenario without loss of generality, see Table~\ref{tab:gqm}.

\begin{table}[ht!]
    \centering
    \small
    \begin{tabular}{|l|l|}
    \hline
    \multicolumn{2}{|c|}{\begin{tabular}[c]{@{}l@{}}\textbf{G1: Improve time to find an eligible adaptation}\\\end{tabular}} \\ 
        \hline
        \multicolumn{1}{|c|}{Questions} & \multicolumn{1}{|c|}{Metrics} \\ 
        \hline
        \begin{tabular}[c]{@{}l@{}}\textbf{Q1}: How much time does the optimization\\ pipeline take to find a solution? \end{tabular} & \begin{tabular}[c]{@{}l@{}} Duration [s] \end{tabular} \\ 
        \hline
    \multicolumn{2}{|c|}{\begin{tabular}[c]{@{}l@{}}\textbf{G2: Improve quality of the adaptation}\\\end{tabular}} \\ 
        \hline
        \multicolumn{1}{|c|}{Question} & \multicolumn{1}{|c|}{Metric} \\ 
        \hline
        \begin{tabular}[c]{@{}l@{}}\textbf{Q2}: How efficient is the proposed learn-\\ing and optimization? \end{tabular} & \begin{tabular}[c]{@{}l@{}}CT metrics\end{tabular} \\
        \hline

    \end{tabular}
    \caption{Goal-Question-Metric }
    \label{tab:gqm}
\end{table}

The first goal (\textbf{G1}) stems from the fact that the proposed AI-based optimization pipeline aims at providing support to the search for an optimal adaptation task by reducing time to solution at runtime. Therefore, to support \textbf{G1}, we collect the duration of the optimization pipeline for both components, Strategy Manager and Strategy Enactor. The optimization pipeline relies on the identification of parameters through the collection of control theoretical metrics from QoS attributes of the system at runtime. Thus, with \textbf{G2}, we want to guarantee that our approximation does lead to efficient solutions. By this means, we measure the efficiency of our method by the reached control theoretical metrics with optimized parameters configuration and the best solutions given with the exhaustive search algorithm.

Under the light of the GQM, 
the experiment is performed in complementary phases (i) time and space evaluation of the search for the optimal adaption, and (ii) evaluation of the adaptation quality with distinct configurations. Thus, following sections further describe the experimental setup, the operational scenario under which the experiments were conducted, and finally the results.

\subsection{Time and space to find an eligible adaptation}
~In this phase, we have first instrumented the pipeline with a chronometer to measure the duration of the learning model step and the finding an optimal configuration step, from both strategy manager instance and strategy enactor instances. 

For the strategy manager, the time to perform the curve fitting (time to solution, gran and offset) was in average 0.3099 seconds per execution, whereas the curve type was classified as exponential. The NGSA-II took in average 0.0023 seconds per execution. Totaling in 0.3123 seconds per execution for the strategy manager pipeline.

For the strategy enactor, four curve fittings had to be realized (i) Kp-SSE, (ii) Kp-Overshoot, (iii) Ki-SSE and (iv) Ki-Overshoot. For (i) the algorithm took 0.01295 seconds to fit the model into a quadratic function, (ii) it took 0.0498 seconds to fit into an exponential, (iii) it took 0.0127 seconds to fit into a quadratic and (iv) it took 0.0075 seconds to fit into a linear model. Furthermore, the NSGA-II was executed twice (v) Kp-SSE \& Kp-Overshoot and (vi) Ki-SSE \& Ki-Overshoot. For (v) the algorithm took 5.3947 seconds and (vi) it took 4.896 seconds. Totaling in 10.374099032 seconds per execution for the strategy manager pipeline.

Our execution of NSGA-II makes ten thousand function evaluations. However, this amount can be reduced to one thousand without great quality loss if the execution time of the pipeline needs to be improved.

\subsection{Quality of the adaptation}

We have deployed and executed the prototype under a scenario where at least one adaptation was triggered for each five minutes execution. Then, 50 randomly selected combinations of $K_p$ and $K_i$, namely configurations, were executed. The 50 configurations are a product of the combination of two sets Kp = [60,150] with steps of 10, Ki = [0.2,1] with steps of 0.2 and IW = 5. Once we select an optimal configuration with the AI technique, another scenario is performed, totaling 51 configuration scenarios. The configuration selected by the optimization is $\textit{Kp} = 114.0$, $\textit{Ki} = 0.75$ and $IW = 5$.

The scenario was derived from the running example, where the stakeholders wish to maintain the system's reliability level at 95\%. The maintenance process is held by the system manager, as follows.

The system manager must continuously monitor the target system's reliability status during runtime. Moreover, the prototype was developed to satisfy complementary tasks with single responsibility components. For that reason, the reliability is locally calculated for each component and composed through a model that relates their probability of success, i.e. the reliability formula from previous works~\cite{solano2019taming}. To keep track of the system goals, the system manager systematically analyzes whether the monitored reliability status fulfills them. Where in case of violations to the adaptation strategy conditions the system managers adapts the target system to maintain the reliability level accordingly to the stipulated goal.

In the experimental sessions, we simulated a scenario where multiple sensors flood the central processor with a higher rate than what it can handle. On top of that, noise is generated by an external agent that triggers random failures on the sensors preventing them from forwarding data to the central processor. The simulated failures reduce the number of messages received by the central processor at a random factor, placing another layer of uncertainty on the system execution. To cope with the disturbances into the reliability status, the system manager delivers reconfiguration signals containing increments/decrements to the sensors' sampling rate that could give more or less time for the central processor to maintain a always flowing processing queue.

\begin{figure}[h]
    \centering
    \centerline{\includegraphics[width=0.5\textwidth]{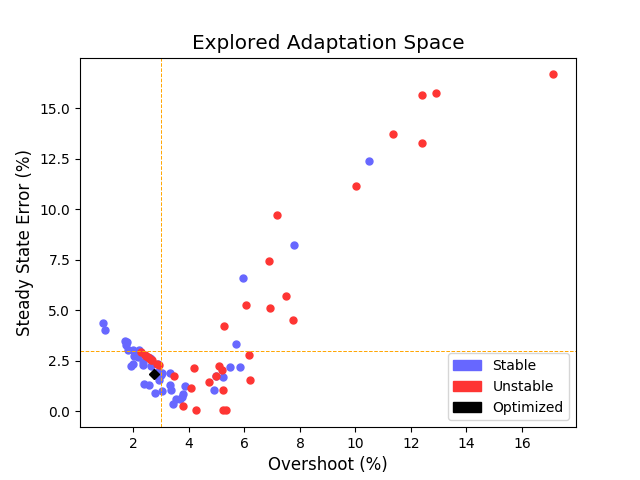}}
    \caption{Explored adaptation space.}
    \label{fig:adaptspace_results}
\end{figure}
\begin{figure}
     \centering
     \begin{subfigure}[b]{0.5\textwidth}
        \centerline{\includegraphics[width=\textwidth]{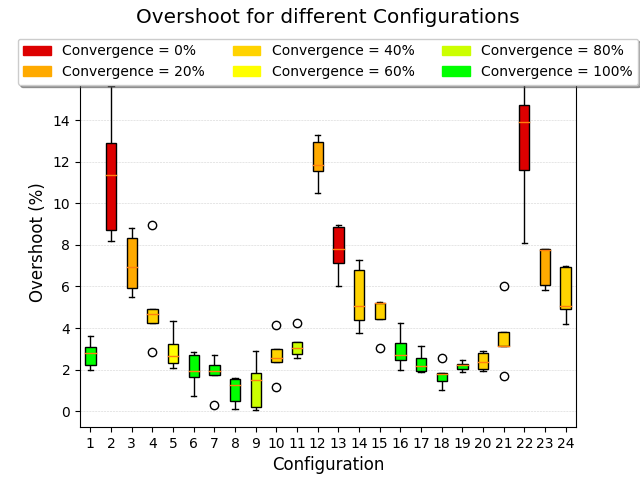}}
        \subcaption{Overshoot for distinct configurations.}
        \label{fig:os_results}
     \end{subfigure}
        
     \centering
     \begin{subfigure}[b]{0.5\textwidth}
        \centerline{\includegraphics[width=\textwidth]{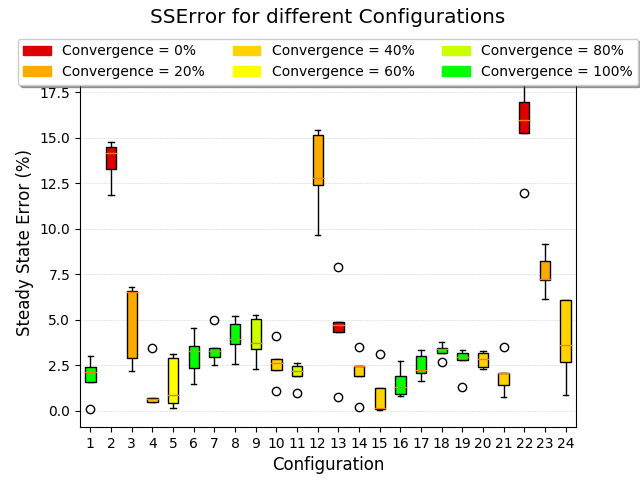}}
        \subcaption{Steady-state-error for distinct configurations.}
        \label{fig:sse_results}
     \end{subfigure}
     
    \caption{Boxplots for the sample of 24 configurations}
    \label{fig:boxplots}
\end{figure}

The SAS was instrumented to log information regarding the quality of the adaptations w.r.t the control theory metrics. The outcoming data is processed accordingly to each evaluation and is presented in Figures \ref{fig:adaptspace_results} and \ref{fig:boxplots}.

Figure~\ref{fig:adaptspace_results} shows the results of the adaptation space explored by the na\"ive (in red and blue) and our approach (in black). Our approach was able to meet the nearly optimal adaptation space for the threshold of steady state error and overshoot of 3\%. Figures \ref{fig:os_results} and \ref{fig:sse_results} provides a bigger picture of the experiments over the 50 configurations and the optimized one. For the sake of space, we refer only to the first 24 configurations, where configuration 1 refers to our approach. Results for the overshoot and steady-state error show that our approach managed to perform among the best results, reaching out convergence point of 100\%. In other words, our approach not only managed to optimize steady-state and overshoot but also the stability of our BSN. By these means, the self-adaptation loop of the BSN from the control theoretic principles can be considered quite robust for the scenarios evaluated.

\subsection{Threats to Validity}

\noindent\textbf{Internal validity.}
The main threat to internal validity is the data that we collected from the scenarios of executing our approach and that we used for the model learning and the optimization. Different scenarios may lead to different data and thus, to different results of our approach. 
Moreover, we rely on specific techniques for the model learning (curve fitting) and optimization (\mbox{NSGA-II}). Other techniques could be used, which may lead to different results. In this context, we have not tuned the techniques, especially the hyper-parameters of NSGA-II, to avoid any bias concerning BSN.

\noindent\textbf{External validity.}
Although we present a generic self-adaptation approach that combines AI and control-theoretical principles, our evaluation focuses on the BSN case study with a single non-functional requirement (reliability) as the adaptation concern. 
Thus, we cannot generalize our results to other concerns (e.g., performance, security, or costs) and target systems, which calls for further experiments.

\noindent\textbf{Construct validity.}
The main threat is the correctness of the implementation of our approach and the BSN used for the evaluation. 
Concerning our approach, at least two authors of this paper have reviewed the implementation and checked the plausibility of the evaluation results based on the experience they have with the BSN from earlier work~\cite{rodrigues2018learning,solano2019taming}. With this experience we are confident about the validity of the BSN that is publicly available. 

\section{Related Work} \label{sec:related}

As stated by \citet{weyns2017software}, the use of control theory in the design of self-adaptive software can bring several benefits since it allows providing analytical guarantees for several system properties such as stability and robustness. 
Consequently, the idea of using control-based approaches to achieve self-adaptation for software systems has been widely studied~\cite{shevtsov:2018}, and the use of control theory as a well-suited solution for self-adaptation to systematically meet the adaptation goals despite a certain degree of uncertainty of the system and environment has been recognized~\cite{filieri2015software,filieri2011self,filieri:2017,diao2005self}.

Control-based self-adaptation approaches typically propose controller synthesis to automatically construct a controller for managing the software system's adaptation needs~\cite{filieri2015automated,shevtsov2016keep,filieri2014automated}. The resulting controllers are related to the Strategy Enactor in our work. These approaches to synthesize these controllers could be applied for the Strategy Enactor whose focus is on using a control-theoretical controller. 
However, our work presents a further module called Strategy Manager that synthesizes an adaptation strategy to which the controller subsequently adheres 
As previously stated, the combination of these two modules is the essence of our work, where self-adaptations is realized at two different levels: we first use a global goal, set by users, to come up with an adaptation strategy, and then we adapt the target system's parameters based on goals, used as setpoints, conditions and actions defined in the strategy. 

Moreover, we use AI-based learning and optimization techniques in our work to reduce the adaptation space at the strategy manager level and to optimize the controller's parameters at the strategy enactor level. For the latter level, the focus is on meeting acceptable system behavior with respect to control properties such as  overshoot and steady-state error. 
In this context, other approaches use learning, online or offline, and other statistical methods to reduce the adaptation space, often called configuration space, in order to find a configuration for the system in available time~\cite{gerostathopoulos2018adapting,Elkhodary:2010,Esfahani:2013,quin:2019,jamshidi:2019,jamshidi2016uncertainty,jamshidi2017transfer}. 
The main difference to our work is that improve efficiency of synthesizing adaptation strategies by optimizing the the search process (in terms of granularity and offset) exploring the adaptation space. Thus, we optimize the exploration of the search space rather than explicitly reducing the space. 

Similarly to the strategy enactor in our work, existing control-based approaches for self-adaptation use PI or PID controllers that are adapted at run-time (cf.~\cite{shevtsov:2018}), to compensate inaccuracies of the model or to handle radical changes. Particularly, values of parameters in the model are estimated and updated at run-time based on measurements often processed by filter algorithms such as Kalman or Recursive Least Square~\cite{Klein:2014,Maggio:2014,Klein:resilience:2014,filieri2014automated,shevtsov2016keep,Shevtsov:2017}, or controller parameters are tuned online by relay feedback~\cite{Filieri:2012} or machine learning~\cite{Lama:2013}, or offline by experiments~\cite{Desmeurs:2015}. In contrast, in our approach the controller parameters are tuned by an evolutionary algorithms identifying (near-)optimal values for them. 
Evolutionary algorithms have been applied previously to self-adaptive systems in order to find optimal configurations of the target system~\cite{Fredericks:2019,Shin:2019} while we apply them to improve the self-adaptation performed by the strategy manager and strategy enactor.

To the best of our knowledge, our work is a novel approach that combines the use of algebraic models (i.e., the parametric formulae) of a system, which are used to synthesize adaptation strategies based on a non-functional requirement, and control theory, which is used to adapt the target system's parameters systematically based on the previously defined strategy. Also, we combine AI-based optimization with these two approaches to adapt the system in an efficient way while also improving the self-adaptation in terms of the control-theoretical properties: overshoot and steady-state~error.

\section{Conclusion and Future Work} \label{sec:conclusion}

In this work, we have proposed a hybrid approach that combines control theory principles with AI techniques to optimize the adaptation process in self-adaptive systems. Using curve fitting methods aligned with a meta-heuristic technique, namely NSGA-II, we were able to i)~efficiently synthesize adaptation strategies through high-level reasoning upon the model that represents the managed system, and (ii)~enforce actions through control-theoretical principles to ensure the adaptation strategies are applied and properties of interest guaranteed. By these means, we contribute with a hierarchical and dynamic system manager that relies on  optimization at all levels of the decision-making process towards a more efficient and robust adaptation mechanism.
We have evaluated our approach on the BSN prototype implemented in the ROS framework. The evaluation results have shown that our hybrid approach is able to find optimal solutions for the adaptation space while also improving the self-adaptation loop in terms of control theoretic properties: stability, overshooting and steady-state error. 

We envision that our future work will be devoted into expanding our technique to accommodate PID controllers as well as exploiting further evolutionary algorithms, while ensuring our approach is also able to optimize the adaptation space at runtime. During the experiments, we have noticed that defining a suitable data window in which the method should train the model is a hard task. A second issue is knowing when the learned model should be applied. Some kind of quality monitor is likely to be incorporated to the architecture in order to orchestrate the learning process. From a performance perspective, this brings us to another challenge, that is building the pipeline in a way that the continuous learning and adaptation of metrics do not affect negatively the operation of the managed system. We plan to further investigate these issues in a future work. Additionally, for generalization purposes, we plan to conduct case studies with other exemplars from other domains than healthcare as well as to further compare our solution to state-of-the-art adaptive control algorithms with respect to time to solution and robustness.

\section*{Acknowledgment}

The authors express their utmost gratitude to L\'eo Moraes and Gabriel Levi (UnB/Brazil) for implementing an accessible version of the BSN for experimentation on SAS. This study was financed in part by the CAPES-Brasil -- Finance Code 001, through CAPES scholarship. 
This work was also partially supported by the Wallenberg Al, Autonomous Systems and Software Program (WASP) funded by the Knut and Alice Wallenberg Foundation, 
the FLASH project (GR~3634/6-1) funded by the German Science Foundation (DFG), 
and the EU H2020 Research and Innovation Prog. under GA No. 731869 (Co4Robots). 
The authors also acknowledge financial support from Centre of EXcellence on Connected, Geo-Localized and Cybersecure Vehicle (EX-Emerge) funded by Italian Government under CIPE resolution n. 70/2017 (Aug. 7, 2017).
Finally, we thank CNPq for partial support under grant number 306017/2018-0.

\balance
\bibliographystyle{ACM-Reference-Format}
\bibliography{reference}

\end{document}